\documentclass[conference]{IEEEtran}
\IEEEoverridecommandlockouts
\usepackage[utf8]{inputenc}
\usepackage{cite}
\usepackage{amsmath,amssymb,amsfonts}
\usepackage{algorithmic}
\usepackage{graphicx}
\usepackage{textcomp}
\usepackage{regensky}

\def\BibTeX{{\rm B\kern-.05em{\sc i\kern-.025em b}\kern-.08em
T\kern-.1667em\lower.7ex\hbox{E}\kern-.125emX}}

\usepackage[firstpage=true]{background}

\SetBgContents{\parbox{\textwidth}{\footnotesize © 2024 IEEE. Personal use of this material is permitted. Permission from IEEE must be obtained for all other uses, in any current or future media, including reprinting/republishing this material for advertising or promotional purposes, creating new collective works, for resale or redistribution to servers or lists, or reuse of any copyrighted component of this work in other works.}}
\SetBgScale{1}
\SetBgAngle{0}
\SetBgPosition{current page.north}
\SetBgVshift{-1.2cm}
\SetBgColor{black}
\SetBgOpacity{1}

\begin{document}

  \title{Analysis of Neural Video Compression Networks for 360-Degree Video Coding
  \thanks{This work was supported by the Deutsche Forschungsgemeinschaft (DFG, German Research Foundation) under project number 418866191.}
  }

  \author{\IEEEauthorblockN{Andy Regensky, Fabian Brand, and André Kaup}
  \IEEEauthorblockA{Multimedia Communications and Signal Processing\\
  Friedrich-Alexander-Universität Erlangen-Nürnberg\\
  Cauerstr. 7, 91058 Erlangen, Germany\\
  \{andy.regensky, fabian.brand, andre.kaup\}@fau.de}
  }

  \maketitle

  \begin{abstract}
      With the increasing efforts of bringing high-quality virtual reality technologies into the market, efficient 360-degree video compression gains in importance.
      As such, the state-of-the-art H.266/VVC video coding standard integrates dedicated tools for 360-degree video, and considerable efforts have been put into designing 360-degree projection formats with improved compression efficiency.
      For the fast-evolving field of neural video compression networks (NVCs), the effects of different 360-degree projection formats on the overall compression performance have not yet been investigated.
      It is thus unclear, whether a resampling from the conventional equirectangular projection~(ERP) to other projection formats yields similar gains for NVCs as for hybrid video codecs, and which formats perform best.
      In this paper, we analyze several generations of NVCs and an extensive set of 360-degree projection formats with respect to their compression performance for \mbox{360-degree} video.
      Based on our analysis, we find that projection format resampling yields significant improvements in compression performance also for NVCs.
      The adjusted cubemap projection (ACP) and equatorial cylindrical projection (ECP) show to perform best and achieve rate savings of more than 55\% compared to ERP based on WS-PSNR for the most recent NVC.
      Remarkably, the observed rate savings are higher than for H.266/VVC, emphasizing the importance of projection format resampling for NVCs.
  \end{abstract}

  \begin{IEEEkeywords}
    learned video compression, 360-degree, video coding, projection format resampling
  \end{IEEEkeywords}

  \section{Introduction}\label{sec:introduction}

The task of video compression is a crucial element of today's multimedia communications landscape.
Among the most prominent video compression standards are hybrid video codecs like the popular H.264/AVC~\cite{Wiegand2003}, its successor H.265/HEVC~\cite{Sullivan2012}, and the state-of-the-art H.266/VVC~\cite{Bross2021a}.

With the growing popularity of augmented and virtual reality systems (AR/VR), the importance of efficient video compression especially of 360-degree video content further increases.
In order to use existing, highly efficient video compression standards, 360-degree videos need to be mapped to the 2D image plane prior to encoding leading to inevitable distortions~\cite{Pearson1990}.
Special efforts in improving compression performance for 360-degree video include the design of improved projection formats~\cite{Ye2020a} and the development of 360-degree specific coding tools~\cite{Sauer2017, Sauer2018, Zhou2020, Vishwanath2022, Regensky2023a}.

In recent years, neural video compression networks (NVCs) have experienced significant advances~\cite{Lu2019, Agustsson2020, Li2021, Sheng2023, Li2022, Li2023} and already perform close to state-of-the-art video compression standards like H.266/VVC that have evolved over decades.
These fast-paced improvements suggest that neural network based video coding technologies will become increasingly important in the future.
However, evaluations of NVC compression performance have so far mainly addressed perspective content.
It is thus unclear, how NVCs behave for 360-degree video in different projection formats and whether specific extensions for 360-degree video are required, as is the case for hybrid video codecs.

\begin{figure}
  \centering
  \tikzstyle{block}=[draw, minimum width=36pt, minimum height=24pt, align=center, node distance=12pt]
\tikzstyle{arrow}=[-latex]
\def\bitstream at (#1,#2){
    \draw[draw=black] (#1 - 0.0625, #2) rectangle (#1 + 0.0625, #2 + 1);
    \fill[fill=black] (#1 - 0.0625, #2 + 0.125) rectangle (#1 + 0.0625, #2 + 0.25);
    \fill[fill=black] (#1 - 0.0625, #2 + 0.375) rectangle (#1 + 0.0625, #2 + 0.5);
    \fill[fill=black] (#1 - 0.0625, #2 + 0.75) rectangle (#1 + 0.0625, #2 + 0.875);
}

\begin{tikzpicture}
  \footnotesize

  \coordinate (bitstream) at (0, 0);
  \bitstream at (0, -0.5);
  \node[block, left=of bitstream] (encoder) {Neural\\Encoder};
  \node[block, right=of bitstream] (decoder) {Neural\\Decoder};

  \node[block, left=30pt of encoder] (fwd_resampling) {Format\\Resampling};
  \node[block, right=30pt of decoder] (bwd_resampling) {Format\\Resampling};

  \node[above=of fwd_resampling] (erp_original) {\small $\vec{x}_\text{erp}$};
  \node[above=of bwd_resampling] (erp_decoded) {\small $\hat{\vec{x}}_\text{erp}$};

  \node[block] at ($(erp_original)!0.5!(erp_decoded)$) (quality_metric) {Quality\\Metric};

  \draw[arrow] (erp_original) -- (fwd_resampling);
  \draw[arrow] (fwd_resampling) -- (encoder) node[midway, above] {$\vec{x}_\text{proj}$};
  \draw[arrow] (encoder) -- ($(bitstream) - (0.052, 0)$);
  \draw[arrow] ($(bitstream) + (0.052, 0)$) -- (decoder);
  \draw[arrow] (decoder) -- (bwd_resampling) node[midway, above] {$\hat{\vec{x}}_\text{proj}$};;
  \draw[arrow] (bwd_resampling) -- (erp_decoded);
  \draw[arrow, dashed] (erp_original) -- (quality_metric);
  \draw[arrow, dashed] (erp_decoded) -- (quality_metric);

  \node[above=16pt of bitstream] {Bitstream};

\end{tikzpicture}
  \caption{Investigated coding framework with the original video $\vec{x}_\text{erp}$ in equirectangular projection and the resampled video $\vec{x}_\text{proj}$ in coding projection. $\hat{\vec{x}}_\text{proj}$ and $\hat{\vec{x}}_\text{erp}$ are the corresponding decoded videos.}
  \label{fig:coding-framework}
\end{figure}
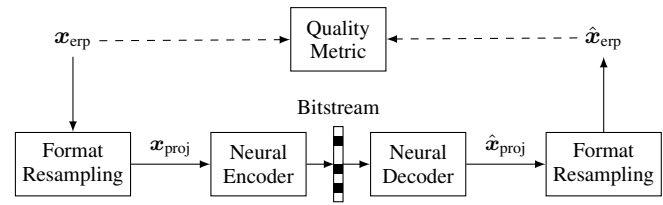

This paper aims to provide clarity on these points by analysing the compression performance of NVCs for 360-degree video.
Similar to the investigations for hybrid video codecs, an extensive set of 360-degree projection formats is tested and compared with respect to their compression performance.
Fig.~\ref{fig:coding-framework} shows an overview of our NVC testing framework that follows the process proposed by the Joint Video Experts Team (JVET) in the common test conditions for 360-degree video~\cite{He2021}.
In addition to showing that projection format resampling becomes even more important for NVCs than it has been for H.266/VVC, we show that its importance is likely to increase even further in future NVCs, based on trends derived from testing multiple generations of NVCs.

  \section{Investigated Networks and\\Projection Formats}\label{sec:state-of-the-art}

\subsection{Deep Contextual Video Compression Networks}\label{subsec:nvcs}

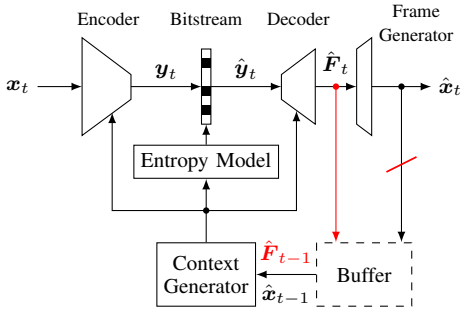
\begin{figure}
    \centering
    \tikzstyle{block}=[draw, minimum width=36pt, minimum height=24pt, align=center, node distance=12pt]
\tikzstyle{thinblock}=[draw, minimum width=36pt, align=center, node distance=12pt]
\tikzstyle{arrow}=[-latex]
\def\bitstream at (#1,#2){
    \draw[draw=black] (#1 - 0.0625, #2) rectangle (#1 + 0.0625, #2 + 1);
    \fill[fill=black] (#1 - 0.0625, #2 + 0.125) rectangle (#1 + 0.0625, #2 + 0.25);
    \fill[fill=black] (#1 - 0.0625, #2 + 0.375) rectangle (#1 + 0.0625, #2 + 0.5);
    \fill[fill=black] (#1 - 0.0625, #2 + 0.75) rectangle (#1 + 0.0625, #2 + 0.875);
}

\begin{tikzpicture}
    \footnotesize
    \bitstream at (0, 0);

    \draw[black] (-1.65, -0.15) -- (-1, 0.33) -- (-1, 0.66) -- (-1.65, 1.15) -- cycle;
    \draw[black] (1.45, 0) -- (1, 0.33) -- (1, 0.66) -- (1.45, 1) -- cycle;
    \draw[black] (2.2, -0.15) -- (2, 0) -- (2, 1) -- (2.2, 1.15) -- cycle;

    \draw[arrow] (-1, 0.5) -- (-0.0625, 0.5) node[midway, anchor=south] {$\vec{y}_t$};
    \draw[arrow] (0.0625, 0.5) -- (1, 0.5) node[midway, anchor=south] {$\hat{\vec{y}}_t$};
    \draw[arrow] (1.45, 0.5) -- (2, 0.5) node[midway] (f_mid) {};
    \node[above=2pt of f_mid.center] {$\hat{\vec{F}}_t$};

    \node at (-2.5, 0.5) (x_t) {$\vec{x}_t$};
    \draw[arrow] (x_t) -- (-1.65, 0.5);

    \node at (3.25, 0.5) (xhat_t) {$\hat{\vec{x}}_t$};
    \draw[arrow] (2.2, 0.5) -- (xhat_t) node[midway] (xhat_mid) {};

    \node[thinblock] at (0, -0.5) (entropy_model) {Entropy Model};

    \node[block] at (0, -2) (context_generator) {Context\\Generator};
    \draw[arrow] (context_generator) -- (entropy_model) node[midway] (cg_mid) {};
    \fill (cg_mid) circle (1.2pt);
    \draw[arrow] (cg_mid.center) -| (-1.25, 0.155);
    \draw[arrow] (cg_mid.center) -| (1.2, 0.18);
    \draw[arrow] (entropy_model) -- (0, 0);

    \coordinate (buffer) at (2.1, -2);
    \node[block, dashed] at (buffer) (b2) {Buffer};
    \fill(xhat_mid) circle (1.2pt);
    \draw[arrow] (xhat_mid.center) -- (xhat_mid |- b2.north) node[midway] (stripe) {};
    \fill[red] (f_mid) circle (1.2pt);
    \draw[arrow, red] (f_mid.center) -- (f_mid |- b2.north);
    \draw[arrow] (b2) -- (context_generator) node[midway] (xf_mid) {};
    \draw[red, line width=0.66pt] ($(stripe) - (0.2, 0.1)$) -- ($(stripe) + (0.2, 0.1)$);
    \node[above=0.5pt of xf_mid.center, red] {$\hat{\vec{F}}_{t-1}$};
    \node[below=1pt of xf_mid.center] {$\hat{\vec{x}}_{t-1}$};

    \node at (-1.3, 1.4) {\scriptsize Encoder};
    \node at (0, 1.4) {\scriptsize Bitstream};
    \node at (1.225, 1.4) {\scriptsize Decoder};
    \node[align=center, anchor=south west] at (2.2, 1) {\scriptsize Frame\\\scriptsize Generator};

\end{tikzpicture}
    \caption{Schematic of DCVC and its extensions. Elements in red show the updated feature buffer introduced with DCVC-TCM. $\vec{x}_t$ describes the input frame, $\vec{y}_t$ the latent space, $\hat{\vec{y}}_t$ the decoded latent space, $\hat{\vec{F}}_{t}$ the last feature of the decoder and $\hat{\vec{x}}_{t}$ the decoded frame at time step $t$.}
    \label{fig:dcvc}
\end{figure}

\vspace{0.33em}\textbf{DCVC:}
The deep contextual video compression network~\cite{Li2021} employs a conditional coding framework.
Hybrid video codecs employ a residual coding framework, where a prediction is generated based on additional side information, subtracted from the current input signal, and only the residual signal is coded.
Conditional coding differs from residual coding by directly providing the given side information to the individual network components allowing them to flexibly learn the relevant features for efficient coding.
Fig.~\ref{fig:dcvc} shows a simplified schematic of the basic DCVC framework.
The buffered frame from the last time step $\hat{\vec{x}}_{t-1}$ is further processed in a context generation network and the learned context is then provided to the individual network components.
Motion estimation and motion compensation are omitted for readability.
Note, however, that the context generator uses motion information to warp the learned features internally.
The entropy model includes a hyperprior model, employs an autoregressive component for spatial correlation, and is conditioned on the learned context for temporal correlation.

\vspace{0.33em}\textbf{DCVC-TCM:}
As an extension to DCVC, DCVC-TCM~\cite{Sheng2023} includes a temporal context mining module and discards the autoregressive component in the entropy model for faster decoding.
As shown in red in Fig.~\ref{fig:dcvc}, instead of using the last decoded frame $\hat{\vec{x}}_{t-1}$ for context generation, the temporal context mining module generates context from the last feature $\hat{\vec{F}}_{t-1}$ before the last convolutional layer of the frame generator.
It performs a multi-scale procedure to generate context at multiple scales.
Motion information is used at the multiple scales to warp learned features accordingly.
The resulting multi-scale temporal contexts are inserted into the encoder, decoder and entropy model networks at the respective scales at different stages of the networks.

\vspace{0.33em}\textbf{DCVC-HEM:}
By improving the entropy model and allowing a multi-granularity quantization, DCVC-HEM~\cite{Li2022} further improves upon DCVC-TCM.
The hybrid spatial-temporal entropy model splits the channels of the latent space and then performs a parallel two-step checkerboard coding.
In the first step, only the even/odd positions of both channel splits are coded based on the available multi-scale temporal contexts, a hyperprior and the last decoded latent space $\hat{\vec{y}}_{t-1}$.
In the second step, the entropy model is additionally conditioned on the spatial information from the first coding step allowing spatial correlation and codes the odd/even positions of both channels.
The multi-granularity quantization allows to adjust rate in a single network without having to train different models for different rate points.
It contains three quantization steps including global quantization based on user input, a fixed learned channel-wise quantization, and a content-adaptive learned spatial-channel-wise quantization.

\vspace{0.33em}\textbf{DCVC-DC:}
Most notable advancements of DCVC-DC~\cite{Li2023} over DCVC-HEM include the improvement of the context generation enabling long-term context and a quadtree-based entropy model.
The modified architecture of the context generation network employs the idea of offset diversity for internal motion compensation, where features are warped at multiple motion vector offsets.
Features warped at the same offset are called a group.
A so-called cross-group fusion network then learns context by combining features from multiple groups in a parallelization friendly manner.
The cross-group fusion network also includes a skip connection throughput of the last feature $\hat{\vec{F}}_{t-1}$, which allows to learn long-term context over multiple frames.
The quadtree entropy coding improves upon the two-step coding method from DCVC-HEM by splitting channels in four parts, and coding spatial positions in a quadtree.
The resulting four-step procedure increases spatial and channel-wise correlation leading to improved coding efficiency.

\subsection{360-Degree Projection Formats}\label{subsec:projections}

\begin{figure}
    \centering
    \scalebox{0.94}{\input{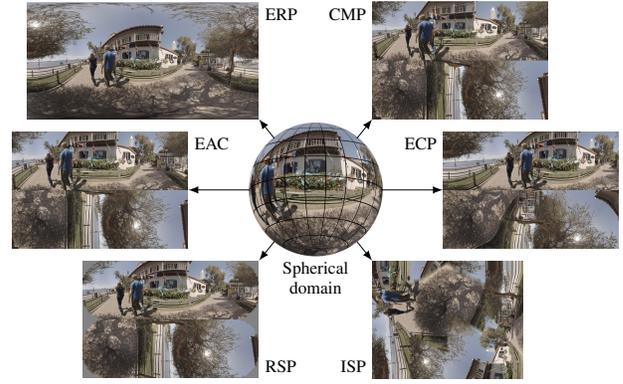}}
    \caption{Overview of a subset of projection formats including the equirectangular (ERP), cubemap (CMP), equiangular cubemap (EAC), equatorial cylindrical (ECP), rotated sphere (RSP), and icosahedron (ISP) projection.}
    \label{fig:projections}
\end{figure}

\vspace{0.33em}\textbf{Equirectangular-based (ERP, AEP):}
The mapping of the classical \textit{equirectangular projection~(ERP)} is identical to the mapping of a classical world map, where the longitudes and latitudes are mapped to the horizontal and vertical image axes, respectively (cf. Fig.~\ref{fig:projections}).
Due to its simplicity, ERP is one of the most common projection formats.
Unlike many projection formats, it entails no discontinuous face boundaries as it consists of a single homogeneous face.
However, this comes at the cost of strong nonlinear distortions especially in the vicinity of the poles.
The \textit{adjusted equalarea projection~(AEP)}~\cite{Zhou2017} is a modification of ERP that alters the sampling rate along the latitudinal direction to better preserve the area of objects in the spherical domain representation.

\vspace{0.33em}\textbf{Cubemap-based (CMP, EAC, HEC, ACP, GCP):}
The \textit{cubemap projection~(CMP)} is derived by projecting the 360-degree image onto the faces of a cube surrounding the image sphere in the spherical domain.
To map a pixel coordinate on the image sphere to a position on the cube, a light ray from the origin to the regarded pixel coordinate is formulated.
The intersection of this light ray with the surrounding cube defines the mapped pixel position.
The six resulting cube faces are arranged onto a 2D plane such that face discontinuities are minimized as much as possible.
As the pixels on the cube faces are arranged in a homogeneous grid, the angular distance between pixel positions on the image sphere decreases with increasing distance to the center of each cube face.
The \textit{equiangular cubemap projection~(EAC)}~\cite{Zhou2017a} ensures consistent angular spacing by warping the cube faces accordingly.
The \textit{hybrid equiangular cubemap projection~(HEC)}~\cite{Lee2018} follows similar goals, but adapts the vertical warping function for cube faces along the equator.
Similarly, the \textit{adjusted cubemap projection~(ACP)}~\cite{Coban2017} and the \textit{generalized cubemap projection~(GCP)}~\cite{Lee2019} use modified face warping functions.

\vspace{0.33em}\textbf{Others (ECP, RSP, ISP):}
The \textit{equatorial cylindrical projection~(ECP)}~\cite{VanDerAuwera2017} yields a single homogeneous face by wrapping a cylinder around the image sphere in longitudinal direction.
Pixel positions up to a specified latitudinal distance from the equator are mapped to this face.
The remaining pixel positions are mapped to the top and bottom caps of the cylinder and warped to fill a rectangular area.
The faces are then split and arranged onto a 2D plane similar to cubemap-based projections.
The \textit{rotated sphere projection~(RSP)}~\cite{Abbas2017a} wraps the image sphere by two segments matching those of a tennis ball, the \textit{icosahedron projection~(ISP)}~\cite{Zakharchenko2016} wraps the image sphere by an icosahedron with 20 faces.
The segments or faces are then arranged space-savingly on the 2D image plane.

  \section{Analytical Methods}\label{sec:analytical-methods}

\subsection{Test Sequences}

For our evaluations, we use the JVET 360-degree test sequences~\cite{He2021}.
The dataset consists of 10 uncompressed sequences in ERP format in yuv color space with 4:2:0 chroma subsampling.
Resolutions range from 6K to 8K with frame rates of 30 to 60 fps and bit depths of 8 and 10 bit.
32 frames of each sequence are tested.

\subsection{360-Degree Projection Format Resampling}

As all test sequences are in ERP format, a projection format resampling is necessary to evaluate the compression performance of the individual projection formats.
We follow the common test conditions for 360-degree video (360-CTC)~\cite{He2021} and use 360Lib-13.1~\cite{360Lib-13.1} for all described operations as used in standardization.
As visualized in Fig.~\ref{fig:coding-framework}, the resampling is performed before encoding and reversed after decoding.
This allows to perform quality metric evaluation in the ERP domain independent of the respective projection format used for coding.
The resampling procedure combines downsampling to roughly 2K, projection format resampling, and conversion to 4:4:4 chroma subsampling.
The downsampled resolution is selected as $2048 \times 1024$ for ERP, AEP, $1800 \times 1200$ for CMP, EAC, HEC, ACP, ECP, RSP, $1816 \times 1232$ for GCP, and $1306 \times 1672$ for ISP.
As noted in the 360-CTC, downsampling ensures that ERP has no unfair advantage against other projection formats.
After decoding, the decoded sequence is resampled back to the original ERP format, resolution and 4:2:0 chroma subsampling.

\subsection{Video Codec Implementations}

For the investigated NVCs described in Section~\ref{subsec:nvcs}, the open source implementations and pretrained network weights~\cite{DCVC2023} published by the original authors are employed.
As we focus on analyzing the compression capabilities using general purpose network weights, a finetuning to 360-degree video is explicitly not performed.
The available network weights have been trained on the Vimeo-90K dataset~\cite{Xue2019a}.
For DCVC and DCVC-TCM, different rate points (rate-distortion-tradeoffs) are achieved through different sets of pretrained network weights.
For DCVC-HEM and DCVC-DC, different rate points are achieved with the same set of pretrained network weights by varying a quality parameter.
We homogenize the two concepts by defining a quality index $q \in \{0, 1, 2, 3\}$ ranging from low- to high-bitrate scenarios.
Depending on the selected quality index $q$ and the regarded NVC, the required set of pretrained network weights are loaded or the according quality parameter is set.
All employed network weights were pretrained on a mean squared error (MSE) distortion loss.
The group of pictures (GOP) is set as $32$.
All networks code in RGB color space.
Color space conversion from YUV 4:4:4 color space to RGB color space and vice versa is performed before and after decoding according to BT.709~\cite{BT709}.

For H.266/VVC, VTM-22.2~\cite{VTM-22.2} is employed in low-delay~(LD) configuration similar to~\cite{Li2023}.
Instead of different quality indices $q$, different rate points are selected through the quantization parameter $\text{QP} \in \{22, 27, 32, 37\}$ ranging from high- to low-bitrate scenarios.
All remaining elements of the testing framework remain unaltered.

\begin{table*}[t]
  \centering
  \caption{Per-codec BD-Rate in \% for the different projection formats with respect to ERP based on YUV-PSNR and YUV-WS-PSNR. Entries marked by $^*$ were calculated with an intersection over union of both rate distortion curves of less than 33\%. Lower is better.}
  \begin{tabular}{l|l||r|r|r|r|r|r|r|r|r|r}
 &  & ERP & AEP & CMP & EAC & HEC & ACP & GCP & ECP & RSP & ISP \\
\hline\multirow[c]{5}{*}{YUV-PSNR} & VTM-22.2 & 0.00\hspace{1.2ex} & -1.70\hspace{1.2ex} & -2.33\hspace{1.2ex} & -25.93\hspace{1.2ex} & -24.85\hspace{1.2ex} & -27.78\hspace{1.2ex} & -26.14\hspace{1.2ex} & -23.17\hspace{1.2ex} & -22.24\hspace{1.2ex} & 82.08\hspace{1.2ex} \\
 \cline{2-12}& DCVC & 0.00\hspace{1.2ex} & 0.48\hspace{1.2ex} & 4.12\hspace{1.2ex} & -14.79\hspace{1.2ex} & -15.44\hspace{1.2ex} & -15.30\hspace{1.2ex} & -11.54\hspace{1.2ex} & -13.61\hspace{1.2ex} & -13.13\hspace{1.2ex} & 43.72\hspace{1.2ex} \\
 & DCVC-TCM & 0.00\hspace{1.2ex} & 0.69\hspace{1.2ex} & 0.84\hspace{1.2ex} & -28.24\hspace{1.2ex} & -27.03\hspace{1.2ex} & -29.05\hspace{1.2ex} & -26.46\hspace{1.2ex} & -25.31\hspace{1.2ex} & -23.59\hspace{1.2ex} & 75.64\hspace{1.2ex} \\
 & DCVC-HEM & 0.00\hspace{1.2ex} & 0.42\hspace{1.2ex} & -4.79\hspace{1.2ex} & -37.37\hspace{1.2ex} & -36.17\hspace{1.2ex} & -38.12\hspace{1.2ex} & -36.19\hspace{1.2ex} & -35.53\hspace{1.2ex} & -31.44\hspace{1.2ex} & 95.36\hspace{1.2ex} \\
 & DCVC-DC & 0.00\hspace{1.2ex} & 0.41\hspace{1.2ex} & -10.32\hspace{1.2ex} & -42.62\textsuperscript{*} & -41.65\textsuperscript{*} & -43.54\textsuperscript{*} & -42.49\textsuperscript{*} & -40.35\hspace{1.2ex} & -36.05\hspace{1.2ex} & 106.51\hspace{1.2ex} \\
\hline& & & & & & & & & & & \\[-7.5pt]\hline\multirow[c]{5}{*}{YUV-WS-PSNR} & VTM-22.2 & 0.00\hspace{1.2ex} & -20.66\hspace{1.2ex} & -22.49\hspace{1.2ex} & -37.84\hspace{1.2ex} & -37.21\hspace{1.2ex} & -39.07\hspace{1.2ex} & -38.16\hspace{1.2ex} & -37.58\hspace{1.2ex} & -34.44\hspace{1.2ex} & -18.31\hspace{1.2ex} \\
 \cline{2-12}& DCVC & 0.00\hspace{1.2ex} & -14.90\hspace{1.2ex} & -13.99\hspace{1.2ex} & -26.41\hspace{1.2ex} & -27.36\hspace{1.2ex} & -26.59\hspace{1.2ex} & -23.93\hspace{1.2ex} & -27.62\hspace{1.2ex} & -24.83\hspace{1.2ex} & -10.01\hspace{1.2ex} \\
 & DCVC-TCM & 0.00\hspace{1.2ex} & -21.87\hspace{1.2ex} & -23.43\hspace{1.2ex} & -42.74\hspace{1.2ex} & -42.23\hspace{1.2ex} & -42.90\hspace{1.2ex} & -41.45\hspace{1.2ex} & -42.94\hspace{1.2ex} & -38.63\hspace{1.2ex} & -21.02\hspace{1.2ex} \\
 & DCVC-HEM & 0.00\hspace{1.2ex} & -28.12\hspace{1.2ex} & -31.70\hspace{1.2ex} & -52.48\textsuperscript{*} & -51.85\textsuperscript{*} & -52.60\textsuperscript{*} & -51.83\textsuperscript{*} & -53.36\textsuperscript{*} & -46.97\textsuperscript{*} & -30.35\hspace{1.2ex} \\
 & DCVC-DC & 0.00\hspace{1.2ex} & -29.66\hspace{1.2ex} & -35.41\hspace{1.2ex} & -56.20\textsuperscript{*} & -55.71\textsuperscript{*} & -56.42\textsuperscript{*} & -56.39\textsuperscript{*} & -56.67\textsuperscript{*} & -50.09\textsuperscript{*} & -33.97\hspace{1.2ex} \\
\end{tabular}

    \label{tab:bd-rate-projections}
\end{table*}

\subsection{Performance Evaluation}

To evaluate the different combinations of 360-degree projection formats and NVCs, the achieved rate distortion performance needs to be assessed.
The rate is obtained as the average bits per pixel of the compressed bitstream.
As distortion metrics, PSNR and WS-PSNR~\cite{Sun2017} are used.
WS-PSNR is an extension of PSNR for 360-degree video that weights the error at each pixel position by the area it covers in the spherical domain.
Both metrics are calculated between the original video~$\vec{x}_\text{erp}$ and the decoded video $\hat{\vec{x}}_\text{erp}$ in ERP format and YUV 4:2:0 color space as shown in Fig.~\ref{fig:coding-framework}.
YUV-PSNR and YUV-WS-PSNR are calculated by weighting the PSNR of the three YUV components as $(w_y, w_u, w_v)=(6,1,1)/8$ matching the evaluation procedure in standardization~\cite{Itu-t2020}.
The achieved rate distortion performance is evaluated using the Bjøntegaard Delta (BD) model~\cite{Bjontegaard2001}.

  \section{Experimental Results}\label{sec:experiments}

For classical hybrid video codecs, significantly improved compression performance can be achieved by resampling 360-degree videos in ERP format to different projection formats for coding.
For NVCs, it has not yet been investigated if projection format resampling yields rate savings as well, how large the potential rate savings are, and how the different projection functions behave compared to the behavior for hybrid video codecs.
In the following, we aim to answer these questions.

Table~\ref{tab:bd-rate-projections} shows the BD-Rate of VTM-22.2 and the described NVCs for the different projection formats.
The anchor is selected as ERP for each codec individually, i.e., the BD-Rates are calculated independently for each row.
The BD-Rates are shown for both YUV-PSNR and YUV-WS-PSNR as quality metrics, which we abbreviate as PSNR and WS-PSNR in the following.
Entries marked by $^*$ were calculated with an intersection over union (IoU) of less than 33\% of both rate distortion curves along the distortion (quality) metric axis and should thus be treated with caution.

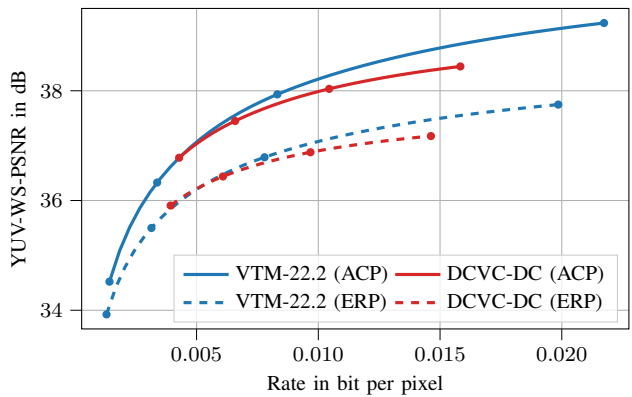
\begin{figure}[t]
  \footnotesize
  \newlength{\axiswidth}
  \newlength{\axisheight}
  \setlength{\axiswidth}{\linewidth}
  \setlength{\axisheight}{0.66\linewidth}
\begin{tikzpicture}

\definecolor{crimson2143940}{RGB}{214,39,40}
\definecolor{darkgray176}{RGB}{176,176,176}
\definecolor{lightgray204}{RGB}{204,204,204}
\definecolor{steelblue31119180}{RGB}{31,119,180}

\begin{axis}[
height=\axisheight,
legend cell align={left},
legend columns=2,
legend style={
  fill opacity=0.8,
  draw opacity=1,
  text opacity=1,
  at={(0.98,0.02)},
  anchor=south east,
  draw=lightgray204
},
tick align=outside,
tick pos=left,
width=\axiswidth,
x grid style={darkgray176},
xlabel={Rate in bit per pixel},
xmajorgrids,
xmin=0.000280153515438239, xmax=0.0227565848040912,
xtick style={color=black},
        xticklabel style={
          /pgf/number format/fixed,
          /pgf/number format/precision=3,
          /pgf/number format/fixed zerofill
        },
        scaled x ticks=false,
y grid style={darkgray176},
ylabel={YUV-WS-PSNR in dB},
ymajorgrids,
ymin=33.6594798125, ymax=39.5006489375,
ytick style={color=black}
]
\addplot [very thick, steelblue31119180]
table {%
0.00142289217975405 34.52112375
0.00183742354009427 35.0820951917133
0.0022519549004345 35.5127499706314
0.00266648626077472 35.859132250681
0.00308101762111495 36.1467993822743
0.00349554898145518 36.3915374875126
0.0039100803417954 36.6061065496445
0.00432461170213563 36.7969938363565
0.00473914306247586 36.9680387914482
0.00515367442281608 37.1222465171189
0.00556820578315631 37.2620091635469
0.00598273714349654 37.3892595423903
0.00639726850383676 37.5055800261485
0.00681179986417699 37.6122813045139
0.00722633122451722 37.7104603891831
0.00764086258485744 37.801044048275
0.00805539394519767 37.8848218224927
0.0084699253055379 37.9625477642043
0.00888445666587812 38.035586501291
0.00929898802621835 38.1045802782014
0.00971351938655858 38.1698926332794
0.0101280507468988 38.2318414347289
0.010542582107239 38.2907061600248
0.0109571134675793 38.3467337854908
0.0113716448279195 38.4001435922607
0.0117861761882597 38.4511311189746
0.0122007075485999 38.4998714364363
0.0126152389089402 38.5465218789031
0.0130297702692804 38.591224336504
0.0134443016296206 38.63410719059
0.0138588329899608 38.6752869565895
0.0142733643503011 38.7148696857295
0.0146878957106413 38.7529521667781
0.0151024270709815 38.7896229610052
0.0155169584313217 38.82496329731
0.015931489791662 38.8590478495277
0.0163460211520022 38.8919454139928
0.0167605525123424 38.923719502286
0.0171750838726827 38.9544288615559
0.0175896152330229 38.9841279327459
0.0180041465933631 39.0128672553822
0.0184186779537033 39.040693826206
0.0188332093140436 39.0676514178016
0.0192477406743838 39.0937808624394
0.019662272034724 39.1191203055765
0.0200768033950642 39.1437054328117
0.0204913347554045 39.1675696735482
0.0209058661157447 39.1907443841642
0.0213203974760849 39.213259013107
0.0217349288364251 39.23514125
};
\addlegendentry{VTM-22.2 (ACP)}
\addplot [semithick, steelblue31119180, mark=*, mark size=1.25, mark options={solid}, only marks, forget plot]
table {%
0.00142289217975405 34.52112375
0.00338176704115338 36.32787125
0.00831569847133425 37.93425875
0.0217349288364251 39.23514125
};
\addplot [very thick, crimson2143940]
table {%
0.00428407357798682 36.77977625
0.00451978859987929 36.8699446724637
0.00475550362177176 36.9539421937091
0.00499121864366423 37.0323679404411
0.0052269336655567 37.105741107801
0.00546264868744917 37.1745144045487
0.00569836370934164 37.2390847995385
0.00593407873123411 37.2998021927026
0.00616979375312658 37.3569764723537
0.00640550877501905 37.4108833052237
0.00664122379691152 37.461783714958
0.00687693881880399 37.5102213001083
0.00711265384069646 37.5564816455639
0.00734836886258893 37.6006910237117
0.0075840838844814 37.6429657360965
0.00781979890637387 37.6834130939695
0.00805551392826634 37.7221322865654
0.00829122895015881 37.7592151512115
0.00852694397205128 37.794746857568
0.00876265899394375 37.8288065166938
0.00899837401583622 37.8614677242246
0.00923408903772869 37.8927990457159
0.00946980405962116 37.922864451144
0.00970551908151363 37.9517237046322
0.0099412341034061 37.9794327146751
0.0101769491252986 38.0060438494529
0.010412664147191 38.03160622123
0.0106483791690835 38.056226729929
0.010884094190976 38.0800470909012
0.0111198092128684 38.1031039499464
0.0113555242347609 38.1254301954289
0.0115912392566534 38.1470567421446
0.0118269542785459 38.1680126835593
0.0120626693004383 38.1883254297511
0.0122983843223308 38.2080208326301
0.0125340993442233 38.2271232998143
0.0127698143661157 38.2456558983665
0.0130055293880082 38.2636404494541
0.0132412444099007 38.2810976148621
0.0134769594317931 38.298046976186
0.0137126744536856 38.3145071074294
0.0139483894755781 38.3304956416548
0.0141841044974706 38.3460293322572
0.014419819519363 38.3611241093734
0.0146555345412555 38.3757951318791
0.014891249563148 38.3900568353804
0.0151269645850404 38.4039229765631
0.0153626796069329 38.417406674224
0.0155983946288254 38.4305204472777
0.0158341096507178 38.44327625
};
\addlegendentry{DCVC-DC (ACP)}
\addplot [semithick, crimson2143940, mark=*, mark size=1.25, mark options={solid}, only marks, forget plot]
table {%
0.00428407357798682 36.77977625
0.00658364858892229 37.44960625
0.0104455325338576 38.03509
0.0158341096507178 38.44327625
};
\addplot [very thick, steelblue31119180, dashed]
table {%
0.00130180948310428 33.9249875
0.0016804739190878 34.4143167247877
0.00205913835507131 34.7853033209674
0.00243780279105483 35.0799046741021
0.00281646722703834 35.3214357035249
0.00319513166302186 35.5241754160688
0.00357379609900537 35.699636901426
0.00395246053498889 35.8545150240294
0.0043311249709724 35.9924023981094
0.00470978940695592 36.1160486838495
0.00508845384293943 36.2276032506577
0.00546711827892295 36.3287766795434
0.00584578271490646 36.4209515649197
0.00622444715088998 36.5052605476507
0.00660311158687349 36.5826425353639
0.00698177602285701 36.6538840241382
0.00736044045884052 36.7196500106173
0.00773910489482404 36.7805074831872
0.00811776933080755 36.8372562066976
0.00849643376679107 36.8907855906193
0.00887509820277458 36.9413898903962
0.0092537626387581 36.9893196400509
0.00963242707474161 37.0347954416367
0.0100110915107251 37.0780125069736
0.0103897559467086 37.1191443725902
0.0107684203826922 37.1583459618482
0.0111470848186757 37.1957561269962
0.0115257492546592 37.2314997734647
0.0119044136906427 37.2656896459992
0.0122830781266262 37.2984278390971
0.0126617425626097 37.3298070811683
0.0130404069985932 37.3599118318161
0.0134190714345768 37.388819223869
0.0137977358705603 37.4165998757258
0.0141764003065438 37.4433185948062
0.0145550647425273 37.4690349891139
0.0149337291785108 37.4938040009082
0.0153123936144943 37.5176763740532
0.0156910580504779 37.5406990646661
0.0160697224864614 37.562915603094
0.0164483869224449 37.5843664139605
0.0168270513584284 37.6050890999546
0.0172057157944119 37.6251186941664
0.0175843802303954 37.6444878850451
0.0179630446663789 37.6632272174546
0.0183417091023625 37.6813652727981
0.018720373538346 37.6989288307644
0.0190990379743295 37.7159430148892
0.019477702410313 37.7324314238306
0.0198563668462965 37.74841625
};
\addlegendentry{VTM-22.2 (ERP)}
\addplot [semithick, steelblue31119180, mark=*, mark size=1.25, mark options={solid}, only marks, forget plot]
table {%
0.00130180948310428 33.9249875
0.00314527187082503 35.49930125
0.00779101931386524 36.78849625
0.0198563668462965 37.74841625
};
\addplot [very thick, crimson2143940, dashed]
table {%
0.00393508109781477 35.90836375
0.0041532346383244 35.9801957717511
0.00437138817883403 36.0469094608321
0.00458954171934366 36.1089961955981
0.00480769525985329 36.1668828863714
0.00502584880036291 36.2209426170425
0.00524400234087254 36.2715031937603
0.00546215588138217 36.3188540735909
0.0056803094218918 36.3632520263904
0.00589846296240143 36.404925796756
0.00611661650291105 36.4440853737382
0.00633477004342068 36.4811814132872
0.00655292358393031 36.5164993651421
0.00677107712443994 36.5501528895841
0.00698923066494957 36.5822456108971
0.0072073842054592 36.6128722448602
0.00742553774596883 36.6421195737485
0.00764369128647845 36.6700672927538
0.00786184482698808 36.696788747509
0.00807999836749771 36.7223515789925
0.00829815190800734 36.7468182893368
0.00851630544851697 36.7702467398306
0.0087344589890266 36.7926905905767
0.00895261252953622 36.8141996897708
0.00917076607004585 36.8348204193327
0.00938891961055548 36.854596002598
0.00960707315106511 36.8735667789332
0.00982522669157474 36.8917989327598
0.0100433802320844 36.9094152023686
0.010261533772594 36.9264487357861
0.0104796873131036 36.9429242151664
0.0106978408536132 36.958864872762
0.0109159943941229 36.9742926002055
0.0111341479346325 36.9892280477798
0.0113523014751421 37.0036907147502
0.0115704550156518 37.0176990317033
0.0117886085561614 37.031270435719
0.012006762096671 37.0444214391099
0.0122249156371806 37.0571676923707
0.0124430691776903 37.0695240419132
0.0126612227181999 37.0815045830905
0.0128793762587095 37.0931227089637
0.0130975297992192 37.1043911552122
0.0133156833397288 37.1153220415452
0.0135338368802384 37.1259269099346
0.013751990420748 37.1362167599572
0.0139701439612577 37.1462020815011
0.0141882975017673 37.1558928850694
0.0144064510422769 37.1652987298873
0.0146246045827866 37.17442875
};
\addlegendentry{DCVC-DC (ERP)}
\addplot [semithick, crimson2143940, mark=*, mark size=1.25, mark options={solid}, only marks, forget plot]
table {%
0.00393508109781477 35.90836375
0.00608230630556742 36.43808125
0.00967854890558455 36.87961375
0.0146246045827866 37.17442875
};
\end{axis}

\end{tikzpicture}
  \caption{Rate distortion performance of VTM-22.2 (blue) and DCVC-DC (red) for ERP (dashed) and ACP (solid) formats.}
  \label{fig:rd-plot}
\end{figure}

Fig.~\ref{fig:rd-plot} provides an example of this behavior showing the rate distortion curves for VTM-22.2 and DCVC-DC for the ERP and ACP formats.
While the curves for VTM-22.2 have sufficient overlap, the curves for DCVC-DC have an IoU of less than 33\% along the distortion axis.
The corresponding BD-Rate value is thus marked by $^*$.
Clearly, the overlap of the curves for DCVC-DC is much larger along the rate axis.
For this reason, all statements based on BD-Rate values marked by $^*$ have been cross-checked using BD-PSNR and BD-WS-PSNR to confirm their validity.

For both BD-Rates based on PSNR and WS-PSNR shown in Table~\ref{tab:bd-rate-projections}, projection format resampling shows significant gains in compression performance compared to coding in ERP format across all codecs.
For our initial investigations, we focus on DCVC-DC as the most recent NVC.
With average rate savings of more than 55\% based on WS-PSNR and more than 40\% based on PSNR, the EAC, HEC, ACP, GCP and ECP projection formats perform best.
Overall, ACP and ECP yield the highest rate savings based on WS-PSNR, though ACP clearly outperforms ECP based on PSNR.
RSP follows closely with rate savings of roughly 50\% based on WS-PSNR and 36\% based on PSNR.
The AEP, CMP and ISP formats cannot compete.
This ranking of projection formats is consistent across all investigated NVC generations.

The EAC, HEC, ACP, GCP and ECP formats have in common that their individual faces exhibit distortions close to those occurring in conventional perspective video, while also being modified to reduce face boundary discontinuities.
If these discontinuities are not adequately taken care of, the compression performance of both VTM-22.2 and the different NVCs suffers noticeably, as visible for the CMP and ISP formats, which show only piecewise perspective characteristics.

It is remarkable that the observed rate savings by applying a reprojection are higher for most NVCs than for VTM-22.2.
For example, while ACP achieves rate savings of more than 56\% based on WS-PSNR for DCVC-DC, rate savings of less than 40\% are achieved for VTM-22.2.
This shows the increased importance of projection format resampling in the context of NVCs.

Investigating the different generations of NVCs, we find that the gap in compression performance between ERP and other formats is steadily increasing, caused by the ongoing specialization of NVCs onto perspective content.
If research on NVCs continues to focus on perspective video, the importance of projection format resampling is thus likely to increase even further in future NVC generations.
However, similar to H.266/VVC that contains specific extensions for 360-degree video, the investigation of specific extensions in NVCs might be a promising approach to further increase compression performance, as well.
Possible approaches could be to finetune network weights for different 360-degree projections, or to investigate 360-degree specific modifications of the internal network architecture.

  \section{Conclusion}\label{sec:conclusion}

In this paper, the compression performance of neural video compression networks (NVCs) is analyzed for their application to 360-degree video coding.
For the evaluation, different generations of NVCs are investigated with respect to their performance on different 360-degree projection formats.
In the context of NVCs, our results show that projection formats like ACP and ECP whose individual faces more closely replicate the characteristics of perspective video, while also considering the reduction of face boundary discontinuities, perform best.
Rate savings of more than 55\% compared to coding in ERP format are observed, showing even higher gains than for H.266/VVC.
Throughout the different NVC generations, the increasing compression performance for perspective content translates into steadily increasing gains of projection formats like ACP and ECP compared to ERP.
Based on the presented results, many promising directions for future research open up.
Possibilities include the optimization of projection format parameters during network training to derive improved projection formats with even higher rate savings, investigating the potential gains if a network is finetuned on 360-degree content in different projection formats, or specifically adapting the network architectures to better handle 360-degree video.

  \bibliographystyle{IEEEbib}
  \bibliography{ms}

\end{document}